\documentclass[twocolumn,showpacs,preprintnumbers,amsmath,amssymb]{revtex4}

\usepackage{graphicx}
\usepackage{dcolumn}
\usepackage{bm}

\begin{document}

\def\bz{\bar z}
\def\R{\mathbb R}
\def\C{\mathbb C}
\def\N{\mathbb N}
\def\Z{\mathbb Z}
\def\A{\mathfrak{A}}
\def\OA{\mathfrak{OA}}
\def\t{\mathfrak t}
\def\x{\mathfrak x}
\def\p{\mathfrak p}
\def\X{\mathfrak X}
\def\P{\mathfrak P}
\def\d{\mathfrak d}
\def\K{\mathfrak K}
\def\J{\mathfrak J}
\def\D{\mathfrak D}
\def\H{\mathfrak H}
\def\I{\hbox{{1\hskip -5.8pt 1}\hskip -3.35pt I}}
\preprint{APS/123-QED}

\title{Extended Non Linear Conformal Symmetry and DSR Velocities on the Physical Surface}

\author{Carlos Leiva}
 \altaffiliation{cleivas@uta.cl}
\affiliation{%
Departamento de F\'{\i}sica, Universidad de Tarapac\'{a} \\
Casilla 7-D, Arica, Chile}%


\date{\today}

\begin{abstract}
The relation between Conformal generators and Magueijo Smolin
Deformed Special Relativity term, added to Lorentz boosts, is
achieved. The same is performed  for Fock Lorentz transformations.
Through a dimensional reduction procedure, it is demonstrated that
a massless relativistic particle living in a $d$ dimensional
space, is isomorphic to one living in a $d+2$ space with pure
Lorentz invariance and to a particle living in a $AdS_{d+1}$
space. To accomplish these identifications, the Conformal Group is
extended and  a nonlinear algebra arises. Finally, because the
relation between momenta and velocities is known, the problem of
position space dynamics is solved.
\end{abstract}

\maketitle
\section{\protect\bigskip Introduction}

 It has been claimed  that the special relativity must be
modified, that a Lorentz symmetry breaking could be observable in
the future in high energy cosmic ray spectra \cite{giovanni,
Stec}, and corrections in the dispersion relation $E^2-p^2=m^2$
has been proposed \cite{piran}. Furthermore, quantum gravity
models suggest that it could be desirable to review the Lorentz
invariance relations and string theories consider some
modifications to the very structure of the space time at high
energy scales. Some data seem to invite to introduce a minimal
length in physical theories, indeed, there are theories that have
some fundamental quantities: the Planck longitude $l_p=\sqrt{\hbar
G/c^3}$, its associated time scale $t_p=l_p/c$ and the Planck
energy $E_p=\hbar /t_p$. It is supposed that beyond these
thresholds, the physics should change dramatically. However, these
absolute values of longitude, time or energy are not in agreement
with the Lorentz transformations, this point reinforces the idea
of to modify the Lorentz boosts.

Solutions for the  problem of how must be, the Lorentz boosts,
modified has been proposed, a very interesting one is Doubly (or
Deformed), special relativity (DSR) theories
\cite{mag,giovanni2,kow}. These theories are based on a
generalization of Lorentz transformations through the more broad
point of view of conformal transformatios, hey have two observer
independent scales, velocity of light and Planck length.

Usually a modification of Lorentz boosts in momentum space is
performed, however when this is done, retrieving the position
space dynamics can be a highly problematic task due to the loss of
linearity. Kimberly, Magueijo and Medeiros \cite{mag2}, have
proposed some methods to achieve this, using a free field theory,
other approaches can be seen in \cite{Wu}, specially the approach
of Deriglazov  is very interesting because he starts from a
conformal group, but it is different from the one proposed here,
because he start at the position space \cite{12}. So, this is a
worth investigating aspect of DSR theories that is not so
understood until today. Other approaches have been done in order
to identify AdS spaces as arenas for DSR theories  and the
approach of this paper can add some information in that way too.

DSR theories are of increasing interest because they can be useful
as a new tool in gravity theories, in Cosmology  as an alternative
to inflation \cite{mof,alb}, and in other fields like propagation
of light \cite{ku}, that is related, for instance, to cosmic
microwave background radiation.

In this paper it is demonstrated that it is possible, formally, to
obtain Fock-Lorentz  \cite{fock,manida}and Magueijo-Smolin
deformations, trough a reduction process and using the conformal
group generators as the generators of the deformed Lorentz
algebra. Then, the open problem of obtaining the space time
dynamics is solved through the relationship on the physical
surface between momenta and velocities on the physical surface
that rise up from the method of dimensional reduction.

More specifically, it is conjectured that the deformations of the
Lorentz algebra in the Fock-Lorentz formulation,  can be treated
as a transformation made by a linear combination of conformal
group generators and the momentum case (the Magueijo Smolin case),
can be understood as the same process, but the inclusion of a new
generator is needed. This new generator can be constructed from
the same theory, and complete the set of symmetries of the
massless Klein Gordon Equation. Then, a DSR massless particle is
shown to be isomorphic to a normal Lorentz particle living in a
$d+2$ space, and the deformations are induced by the dimensional
reduction.

\newpage


\section{\protect\bigskip Conformal Symmetry of Massless
Particle}

The infinitesimal transformations
\begin{equation}
\delta x^\mu = \omega^\mu{}_\nu x^\nu + \alpha^\mu + \beta x^\mu +
2(x\gamma)x^\mu - x^2\gamma^\mu \label{inf1}
\end{equation}
generate the conformal symmetry, $ds^2\rightarrow
ds'{}^2=e^{2\sigma}ds^2$, on the $d$-dimensional Minkowski space
$\R^{1,d-1}$ with metric
$$
ds^2=dx^\mu dx^\nu\eta_{\mu\nu}=-dx_0^2+\sum_{i=1}^{d-1}dx^{ 2}_i.
$$
Here the parameters $\omega^\mu{}_\nu$, $\alpha^\mu$, $\beta$ and
$\gamma^\mu$ correspond to the Lorentz rotations, space-time
translations, scale and special conformal transformations. Due to
a nonlinear (quadratic) in $x^\mu$ nature of the two  last terms
in (\ref{inf1}), the finite version of the special conformal
transformations,
\begin{equation}
x'{}^\mu=\frac{x^\mu-\alpha^\mu x^2}{1 -2\alpha x + \alpha^2 x^2},
\label{confin}
\end{equation}
is not defined globally, and to be well defined requires a
compactification of $\R^{1,d-1}$ by including the points at
infinity.

On the classical phase space with canonical Poisson bracket
relations $\{x_\mu,$ $p_\nu\}=\eta_{\mu\nu}$,
$\{x_\mu,x_\nu\}=\{p_\mu,p_\nu\}=0$, the transformations
(\ref{inf1}) are generated by

\begin{eqnarray}
M_{\mu \nu } = x_{\mu }p_{\nu }-x_{\nu }p_{ \mu },\nonumber\\
P_{\mu }= p_{\mu },\qquad D = x^{\mu }p_{\mu }, \nonumber\\
 K_{\mu } = 2x_{\mu }(xp)-x^{2}p_{\mu }. \label{genconf}
\end{eqnarray}
The generators (\ref{genconf}) form the conformal algebra

\begin{eqnarray}
\{M_{\mu\nu},M_{\sigma\lambda}\}&=&
\eta_{\mu\sigma}M_{\nu\lambda}- \eta_{\nu\sigma}M_{\mu\lambda}
\nonumber\\
& &+\eta_{\mu\lambda}M_{\sigma\nu}-
\eta_{\nu\lambda}M_{\sigma\mu},\nonumber\\
\{M_{\mu \nu },P_{\lambda }\} &=& \eta_{\mu \lambda }P_{ \nu }-
\eta_{\nu\lambda }P_{\mu },\nonumber\\
\{M_{\mu \nu },K_{\lambda \}} &=& \eta_{\mu \lambda }K_{ \nu }-
\eta_{\nu\lambda }K_{\mu },\nonumber\\
\{D, P_{\mu }\} &=&P_{\mu },\nonumber\\
\{D, K_{\mu }\} &=&-K_{\mu },\nonumber\\
\{ K_{\mu },P_{\nu }\} &=& 2(\eta_{\mu \nu }D+M_{\mu
\nu}),\nonumber \\
\{D, M_{\mu\nu}\} &=& \{P_\mu,P_\nu\} =0, \nonumber \\
\{K_{\mu },K_{\nu }\} &=& 0 \label{confalg}
\end{eqnarray}
The algebra (\ref{confalg}) is isomorphic to the algebra
$so(2,d)$, and by defining
\begin{eqnarray}
J_{\mu \nu }=M_{\mu \nu },\quad
 J_{\mu d}=\frac{1}{2}(P_\mu + K_\mu),\nonumber \\
 J_{\mu (d+1)}=\frac{1}{2}(P_\mu-K_\mu),
 \quad
 J_{d(d+1)}=D,
 \label{ident}
\end{eqnarray}
can be put in the standard form
\begin{eqnarray}
\{J_{AB},J_{LN}\}= \eta_{AL}J_{BN}- \eta_{BL}J_{AN} \nonumber \\
+\eta_{AN}J_{LB}- \eta_{BN}J_{LA} \label{so24}
\end{eqnarray}
with $A,B=0,1,d,d+1$, and
\begin{equation}
\eta_{AB}=diag (-1,+1,\ldots,+1,-1). \label{etaab}
\end{equation}

The phase space constraint $ \varphi_m\equiv p^2+m^2=0 $
describing the free relativistic particle of mass $m$ in
$\R^{1,d-1}$ is invariant under the Poincar\'e transformations,
$\{M_{\mu\nu},\varphi_m\}= \{P_\mu,\varphi_m\}=0$. Unlike the
$P_\mu$ and $M_{\mu\nu}$, the generators of the scale and special
conformal transformations commute weakly with $\varphi_m$ only in
the massless case $m=0$: $\{D,\varphi_0\}=2\varphi_0=0$,
$\{K_\mu,\varphi_0\}=4x_\mu\varphi_0=0$.

\section{Dimensional Reduction}

 It was demonstrated in Leiva and Plyushchay \cite{LP},  that a dimensional reduction process from
a massless particle living in $(d+2)$ dimensional space
$\R^{2,d}$,  with  coordinates $H^A$ and metric (\ref{etaab}), and
introducing  canonical momenta $\Pi_A$,
($\{H_A,\Pi_B\}=\eta_{AB}$) can transform  the $so(2,d)$
generators
\begin{equation}
\J_{AB}=H_A \Pi_B - H_B \Pi_A. \label{jxp},
\end{equation}
into the conformal generators in $(d)$-dimensional space
$\R^{1,d-1}$. In order to achieve this, three constrains are added
to the theory:

\begin{eqnarray}
&\phi_0\equiv \Pi_A\Pi^A=0,&
\label{xp2}\\
&\phi_1\equiv H^{A} H_{A}=0,\qquad \phi_2\equiv
H^{A}\Pi_A=0.&\label{xp1}
\end{eqnarray}

The process is implemented trough suitable canonical
transformations and then a dimensional reduction is performed
 onto  the surface defined by $\phi_1$
and $\phi_2$. So in terms of new $\R^{2,d}$ canonical variables
$\tilde H^\mu$, $\tilde\Pi_\mu$ with  $\{\tilde H^\mu,
\tilde\Pi_\mu\}=1$ as defined in \cite{LP}, the $so(2,d)$
generators are the following:

\begin{eqnarray}
\J_{\mu\nu}&=&\tilde H_\mu\tilde\Pi_\nu- \tilde H_\nu\tilde\Pi_\mu,\notag \\
\J_{\mu +}&=&\tilde\Pi_\mu,\notag \\
\J_{d(d+1)}&=&\tilde H_\mu\tilde\Pi^\mu+
2\frac{\tilde\Pi_+}{\tilde H^-}\phi_1 -\phi_2, \nonumber\\
\J_{\mu -}&=&2(\tilde H_\nu\tilde\Pi^\nu)\tilde H_\mu
-(\tilde\Pi_\nu\tilde H^\nu)\tilde\Pi_\mu
\nonumber\\
& &+ \frac{1}{\tilde H ^{-2}}(\tilde\Pi_\mu +4\tilde\Pi_+\tilde
H^-\tilde H_\mu)\phi_1 -2\tilde H_\mu\phi_2.\notag \label{jotas}
\end{eqnarray}

To execute  this process it is supposed that $\tilde H^- \neq 0$.
The constraints now, look like:
\begin{eqnarray}
&\tilde\Pi_\mu\tilde\Pi^\mu=0,&\label{tp2}
\\
&\tilde H^+=0,\qquad \tilde\Pi_-=0.&\label{x+p-}
\end{eqnarray}

The variables $\tilde\Pi_\mu$ and $\tilde H_\mu$ are observable
(gauge invariant) variables with respect to the constraints
(\ref{tp2}) and (\ref{x+p-}), whereas $\tilde H^-= H^-$ and
$\tilde\Pi_+=\Pi_+$ are not. They can be removed by introducing
the constraints
\begin{equation}
\phi_3\equiv H^-+1=0,\quad \phi_4\equiv\Pi_+=0 \label{gaugex}
\end{equation}
as gauge fixings of (\ref{x+p-}) and the reduction to the physical
surface achieves the conformal generators in $\R^{1,d-1}$ as
projections of the $so(2,d)$ generators. Now, with the
identification $x^\mu=\tilde H^\mu$ and $p_\mu=\tilde\Pi_\mu$ and
(\ref{tp2}),  the original massless relativistic particle is
totally retrieved.

\section{Extended Non-linear Conformal Algebra}

It is possible, trough the procedure described  in the last
section, and because positions and momenta are at same level in
the hamiltonian formulation,  to introduce a new generator
$\tilde{K}_{\mu }$:

\begin{equation}
\tilde{K}_\mu = 2p_\mu (xp)-p^2 x_\mu \label{ktilde}
\end{equation}

To do this, we can perform a canonical transformation in
(\ref{jotas}), $\tilde{H}^{A} \rightarrow -\breve{\Pi}^A$ and
$\tilde{\Pi}^{A} \rightarrow \breve{H}^A$ and now to identify
$x^\mu=\breve H_\mu$ and $p_\mu=\breve\Pi_\mu$ . Doing this,
$\J_{\mu -}$ is projected as $\tilde{K}$, and $\J_{\mu +}$ is
projected as $x^\mu$, but the particle  has now the constrain
$x^2=0$. The nature of this generator is totally new because it
produces conformal transformations on the momentum space and the
finite version of this transformations is :

\begin{equation}
p'{}^\mu=-\frac{p^\mu-\alpha^\mu p^2}{1 -2\alpha p + \alpha^2
p^2}, \label{ptrans}
\end{equation}

This is is a symmetry of the massless Klein-Gordon equation too,
in fact: $\{\tilde{K},P^2\}=2P^2 P_\mu=0$. So, after the inclusion
of $\tilde{K}$, aside the relations (\ref{confalg}) we have:

\begin{eqnarray}
\{M_{\mu \nu },\tilde{K}_{\lambda }\} &=& \eta_{\mu \lambda }\tilde{K}_{ \nu}- \eta_{\nu \lambda }\tilde{K}_{\mu },\notag\\
 \{D, \tilde{K}_{\mu }\} &=&\tilde{K}_{\mu }, \notag\\
\{ \tilde{K}_{\mu },P_{\nu }\} &=& 2P_\mu P_\nu- P^2 \eta_{\mu \nu},\notag \\
\{ \tilde{K}_{\mu },\tilde{K}_{\nu }\} &=&0, \notag \\
\{ \tilde{K}_{\mu },K_{\nu }\} &=& 4(M_{\mu \nu}-D\eta_{\mu
\nu})D-2P_\mu K_\nu \nonumber \\ & &  +\frac{2(2DP_\mu
-\tilde{K}_\mu)(2DP_\nu-\tilde{K}_\nu)}{\tilde{K}P}
\label{confalgdef}
\end{eqnarray}

Due to the last  bracket, the set of generators form a non linear
algebra.

We can then, to have two separate set of symmetries. Each set form
a dynamic $SO(2,1)$ symmetry of the Klein Gordon equation,: one
that includes $K_\mu$, generating special conformal
transformations in the position space. And the other that includes
$\tilde{K}_\mu$, generating special conformal transformations in
the momentum space.

The existence of $\tilde{K_\mu}$ could produce some consequences
in the symmetry group of non relativistic particles since there is
a canonical relation between a non relativistic particle in
$\mathbb{R}^1$ and  the relativistic one in $\mathbb{R}^{1,1}$
\cite{LP}. This can be of some interest because Fluid Mechanics
symmetries are intimately related to the free non relativistic
particle symmetry \cite{LO}.

Each group of $SO(2,1)$ symmetries can be used to construct the
DSR transformations in position or momentum space. In the two next
sections,  both cases  shall be reviewed

\section{Fock-Lorentz transformations}

The Fock-Lorentz transformation \cite{fock} \cite{manida}, are
introduced as general linear-fractional transformations between
spatial and temporal coordinates and the result is leaded by
symmetry arguments to the final form:

\begin{eqnarray}
 t'= \frac{\gamma (u)
(t-\mathbf{u}\mathbf{r}/c^2)}{1-(\gamma(u)-1)ct/R
+\gamma(u)\mathbf{u}\mathbf{r}/Rc},\\
\mathbf{r}\|=\frac{\gamma (u) (\mathbf{r}\|
-\mathbf{u}t)}{1-(\gamma(u)-1)ct/R
+\gamma(u)\mathbf{u}\mathbf{r}/Rc},\\
\mathbf{r}\bot=\frac{\mathbf{r}\bot}{1-(\gamma(u)-1)ct/R
+\gamma(u)\mathbf{u}\mathbf{r}/Rc},\\
\end{eqnarray}
where $\mathbf{r}\|$ is the component of the position vector
parallel to the boost $\mathbf{u}$ and $\mathbf{r}\bot$ is the
perpendicular one. Where  $\gamma$  is the Lorentz factor
$\sqrt{1-u^2/c^2}$ and $R$ is a constant with the dimension of
length.

These transformations can be seen as Lorentz transformations for
the quantities:

\begin{equation}
\tilde{t}= \frac{t}{1+ct/R},\qquad
\mathbf{\tilde{r}}=\frac{\mathbf{r}}{1+ct/R} \label{trtilde}
\end{equation}

They can be treated as results of a infinitesimal transformations:

\begin{equation}
\delta x^\mu= \alpha^\nu \{x^\mu ,2x_\nu (xp)\} \label{trapos}
\end{equation}
where $x^0=ct$.

In order to obtain this generator, it can be represented as the
projection of
\begin{equation}
(\J_{\mu -} + \J_{\mu +}\J_{d (d+1)}), \label{jnolin}
\end{equation}
 in $R^{d+2}$ space, wich is
a nonlinear combination of the isometries there. However $2x_\nu
(xp)$ is not a symmetry of the Klein-Gordon equation because
$\{2x_\nu (xp), \varphi_0\}=4p_\mu (xp) + 4x_\mu \varphi_0 \neq
0$. So, it is not possible  to include it into  the generators set
of the massless relativistic particle symmetries.

An alternative, and linear solution is to see (\ref{trtilde}) as
  (\ref{confin}) reduced to  the surface
$x^2=0$ and with the choice $\alpha=(-1/2R,0,0,0)$ and identifying
$x^0$ with $ct$. In this sense, the Fock Lorentz transformations
can be seen as a linear combination of conformal generators $$
M_{0 \mu } + K_{ \mu },$$
 that is a projection of
$$(\J_{\mu \nu} + \J_{\mu -})$$
(which is linear too), for a particle living on the cone (light
cone) $x_{1}^{2}+x_2^2+x_3^2=x_0^2$.

Then we can construct the Fock-Lorentz transformations for a
particle living on the light cone through the  projection of the
action of a linear combination of conformal generators.

Then, Fock Lorentz appear to be deformations of the usual Lorentz
isometries induced by a projection of the system on the physical
surface.
\section{The Magueijo Smolin DSR momentum transformation}

 On the other hand, the Lorentz boost proposed in Magueijo and Smolin \cite{mag}

\begin{equation}
K^i= L_{0 i} +l_{p} p^{i} p_{\mu} \frac{\partial} { \partial
p_{\mu}} \label{lomod}
\end{equation}
where $l_p$ is the Planck length,  can be exponentiated  as

\begin{equation}
K^i= U^{-1}(p_0)L_{0 i}U(p_0)
\end{equation}
where $U(p_0)=exp(l_p p_0 p_{\mu} \frac{\partial}{ \partial
p_{\mu}})$, and the action of $U(p_0)$ over $p_\mu$ is

\begin{equation}
U(p_0)p_\mu= \frac{p_\mu}{1-l_p p_0} \label{p0}
\end{equation}

The generator added to the Lorentz boost in (\ref{lomod}) is the
operational representation of $p^{i}(xp)$ and it can be  as
$\tilde{K}_\mu$, that  produces the following transformation :

\begin{equation}
p'{}^\mu=-\frac{p^\mu-\alpha^\mu p^2}{1 -2\alpha p + \alpha^2
p^2}, \label{confinp}
\end{equation}
to achieve the identification with (\ref{p0}), the transformation
parameter must be $\alpha^\mu= (-l_p/2,0,0,0)$ and the global
minus sign must be removed.

It is possible then, to prescribe  how to obtain the position and
momenta transformation using conformal generators with some
$\tilde{K}$ added. Where does this new generator come from? Some
clues will be seen in the next section.


\bigskip
\bigskip

\bigskip

Starting with the ordinary Lorentz generators:

\begin{equation}
L_{ab}=p_a \frac{\partial }{\partial p^b} - p_b
\frac{\partial}{\partial p^a}
\end{equation}

The modified Lorentz boost proposed in Magueijo and Smolin
\cite{mag} is:

\begin{equation}
K_i= L_{0 i} +l_{p} p_{i} p_{\mu} \frac{\partial} { \partial
p_{\mu}}, \label{lomod}
\end{equation}
where the second term is the deformation proposed and  $l_p$ is
the Planck length. It can be exponentiated as:

\begin{equation}
K^i= U^{-1}(p_0)L_{0 i}U(p_0)
\end{equation}
where $U(p_0)=exp(l_p p_0 p_{\mu} \frac{\partial}{ \partial
p_{\mu}})$, and the action of $U(p_0)$ over $p_\mu$ is

\begin{equation}
U(p_0)p_\mu= \frac{p_\mu}{1-l_p p_0} \label{p0}
\end{equation}

Following the Magueijo Smolin procedure, it can be seen that
boosts in the z direction (as an example), are :

\begin{eqnarray}
p'_0&=&\frac{\gamma (p_0-vp_z)}{1+l_p (\gamma -1)p_0-l_p \gamma
vp_z}\notag \\
p'_z&=&\frac{\gamma (p_z-vp_0)}{1+l_p (\gamma -1)p_0-l_p \gamma
vp_z}\notag \\
p'_x&=&\frac{p_x)}{1+l_p (\gamma -1)p_0-l_p \gamma
vp_z}\notag \\
p'_y&=&\frac{p_y)}{1+l_p (\gamma -1)p_0-l_p \gamma vp_z}.
\end{eqnarray}

This transformations are identical to those obtained by Fock
\cite{fock,manida}, but applied to momentum space. They can be
retrieved replacing p by x and $p_0$ by $t$, instead.

The Fock like  transformations can also be obtained as usual
Lorentz transformations for the transformed  $p'$ from eq \ref{p0}
 \cite{fock,manida}, so we need lead just with the extra term of the deformed
boosts.

 The transformation  (\ref{p0}) can be yielded as result  of the
action of an  $\tilde{K}_\mu = 2p_\mu (xp)-p^2 x_\mu$ generator,
that produces:

\begin{equation}
p'{}^\mu=-\frac{p^\mu-\alpha^\mu p^2}{1 -2\alpha p + \alpha^2
p^2}. \label{confinp}
\end{equation}

So, the Magueijo Smolin transformation can be seen as a
combination of $$ -M_{0 \mu } + \tilde{K}_{ \mu },$$
 that is a projection of
$$(\J_{\mu \nu} + \J_{\mu -})$$
as projected from the $R^{d+2}$ to the particle living on the
light cone, $x_{1}^{2}+x_2^2+x_3^2=x_0^2$, this is the same
condition we had before, and it is a condition for a truly
massless relativistic particle.

Then we can construct the Magueijo Smolin transformations for a
particle living on the light cone through the  projection of the
action of a linear combination of conformal generators. But we
must consider a massless particle, then  we can achieve the
identification with (\ref{p0}), in order to do that it is
necessary to  transform the constrain $\psi_1=p^2\approx 0$ in an
exact identity, the transformation parameter must be $\alpha^\mu=
(l_p/2,0,0,0)$ and the global minus sign must be removed.

\bigskip

\section{The projection onto the physical surface.}

The fundamental problem with the momentum space formulation is to
retrieve the position space. This is a task that has called the
attention of many researches \cite{mag2,Wu,12}. The advantage of
the point of view in the way that is managed here, is that the
task is replaced by the known problem of relating  momenta and
velocities when a dimensional reduction is performed. Just as an
example, let's recall the process to project a free massless
relativistic particle onto the physical surface. This particle has
the Lagrangian:

\begin{equation}
L=\frac{\dot{x}^2}{2e}
\end{equation}
the constrain that rules the dynamic of the particle are:

\begin{eqnarray}
\varphi_e=p_e \approx 0\\
 \varphi_1 =\frac{1}{2}(p^2)\approx 0 \label{const1}
\end{eqnarray}
The condition over the conjugated momentum of $e$, $p_e=0$ shows
that $e$ is a non dynamic variable and we will not take care of it
in the following.

Due the constrain $\varphi_1$, the relation between momenta in
phase space and velocities in the configuration space is not so
clear as it would be desirable, because all momenta yielding
Eq.~(\ref{const1}) should be identified. This ambiguity can be
seen as produced by the fact that the system is constrained to a
subspace in the phase space, this subspace is usually named as the
physical surface.

In order to obtain a truly relationship between momenta and
velocities a reduction of the system to this physical surface must
be performed, this can be done by the following process:

First, we can introduce a gauge fixing to the constrain
Eq.~(\ref{const1}), a suitable one is:

\begin{equation}
\varphi_2=x^0-\tau \approx 0
\end{equation}

Now, it is necessary to reformulate the parenthesis definition
abandoning the Poisson definition and using the Dirac Parenthesis
instead. It is defined by the relation:

\begin{equation}
\{A,B\}_D=\{A,B\}_P-\{A,\varphi_{i} \}_{P} C^{-1}_{ij}\{\varphi_
{j},B\}_{P},
\end{equation}
where $\{\}_P$ indicates Poisson bracket and $\{\}_D$ Dirac
bracket. The matrix C is defined by:

\begin{equation}
C_{ij}=\{\varphi_i,\varphi_j\}_P
\end{equation}
in this case:
\begin{equation}
C_{i,j}= p_0 \left(
\begin{array}{cc}
0 & -1 \\
1 & 0
\end{array}
\right) , \label{2.3a}
\end{equation}
\bigskip

\noindent with

\begin{equation}
C_{i,j}^{-1}= \frac{1}{p_0}\left(
\begin{array}{cc}
0 & 1 \\
-1 & 0
\end{array}
\right), \label{2.3b}
\end{equation}

So, the Dirac brackets are:

\begin{eqnarray}
\{x^0,p_0\}_D&=&0\\
\{x^0,p_i\}_D&=&\frac{p_i}{p^0}\\
\{x^i,p^j\}_D&=&\delta^{ij}
\end{eqnarray}

Now, in order to retrieve the Hamiltonian relations, we define the
independent variables:

\begin{equation}
u^a=(x^i,p_i)
\end{equation}
and introduce the following parameterization:

\begin{equation}
z^s(u,\tau)=(x^0=\tau ,x^i, p^0=+\sqrt{p_i^2}, p^i)
\end{equation}

It is, now, necessary to redefine our Hamiltonian, this can be
performed setting:

$$\mathcal{H}=H+F$$
where $H$ is the former canonical Hamiltonian and F is a
correction due to the projection of the system to the physical
surface, $F$ yields the following equation:

\begin{equation}
\frac{\partial F}{\partial u^m}=-\frac{\partial z^s}{\partial u^m}
W_{st}\frac{\partial z^t}{\partial \tau}
\end{equation}
where $W_{st}$ is defined by:

$$W_{st}=\{x^\mu ,p_\nu\}$$

\begin{equation}
W_{st}= \left(
\begin{array}{cc}
0 & -\delta^\mu_\nu \\
\delta^\mu_\nu & 0
\end{array}
\right), \label{W}
\end{equation}
after proper calculations we have

\begin{eqnarray}
F&=&\sqrt{\vec{p}^2}\\
H&=&0 \\
\mathcal{H}&=&\sqrt{\vec{p}^2}
\end{eqnarray}
and the velocity induced on the physical surface is:

\begin{equation}
\dot{x}^{i}(\tau)=\{x^i,\mathcal{H}\}_D=\frac{p^{i}}{\sqrt{\vec{p}^2
}}
\end{equation}

Just now, we are sure what the relation between velocities and
momenta is, on the physical surface and we can perform the
transformation (\ref{confinp}):

\begin{equation}
\bar{p}_i=\frac{p_i}{1-2l_{p}\sqrt{p_i^2}}
\end{equation}

and, identifying $\sqrt{p_i^2}$ with the energy $E$, the velocity
expression yields:

\begin{eqnarray}
\dot{x}^{i}=\frac{p^{i}}{E} \nonumber\\
 \dot{\bar{x}}^{i}=\frac{p^{i}}{E-2l_{p} E^2}
\end{eqnarray}

Now, in order to obtain the complete Magueijo Smolin velocity we
just need to perform an usual Lorentz boost, for example in $z$
direction:

\begin{eqnarray}
v_x=\frac{(1-u^2)P_x}{(E-2l_{p}E^2-uP_z)}\nonumber\\
v_y=\frac{(1-u^2)P_y}{(E-2l_{p}E^2-uP_z)}\nonumber\\
v_x=\frac{P_{z}-u(E-2l_{p}E^2)}{(E-2l_{p}E^2-uP_z)}\nonumber\\
\end{eqnarray}

\section{DSR and AdS Spaces}
The $(d+2)$-dimensional system (\ref{xp2}), (\ref{xp1}) can be
reinterpreted as a massless particle living on the border of the
AdS${}_{d+1}$ of infinite radius, whose isometry corresponds to
the conformal symmetry of the massless particle in $d$ dimensions.
This can be done in the following way. The massless particle on
AdS${}_{d+1}$ of radius $R$ can be described by the constraints
\begin{equation}
\phi_0=\P_A \P^A=0,\qquad \phi_1=\X_A\X^A+R^2=0, \label{adsl}
\end{equation}
where $\X^A$ and $\P_A$ are the canonical variables. The Poisson
brackets of the constraints (\ref{adsl}) are $\{\phi_1,\phi_0\}=
4\X^A\P_A$. To have the reparametrization-invariant system, the $
\phi_0$ has to be the first class constraint. This can be achieved
by postulating the constraint
\begin{equation}
\phi_2=\X^A\P_A=0 \label{scale}
\end{equation}
in addition to the constraints (\ref{adsl}). The constraint
(\ref{scale}) generates the local scale transformations which due
to the relation $\{\phi_2, \phi_0\}=2\phi_0$ are consistent with
the reparametrization invariance generated by the constraint
$\phi_0$. Since $\{\phi_2, \phi_1\}=-\phi_1+R^2\neq0$, the
constraint $\phi_1=0$ can be understood as a gauge condition for
the constraint (\ref{scale}). Now, let us realize a canonical
transformation $(\X^A,\P_A) \rightarrow (X^A,P_A)$,
\begin{equation}
X^A=\frac{\X^A}{R^{1+\varepsilon}}, \quad
P_A=\P_AR^{1+\varepsilon}, \label{trick}
\end{equation}
with a constant $\varepsilon>0$, and take a limit
$R\rightarrow\infty$, $\X^A\rightarrow\infty$, $\P_A\rightarrow 0$
in such a way that the variables (\ref{trick}) would be finite.
Then, the constraints (\ref{adsl}), (\ref{scale}) take the form of
the first class constraints (\ref{xp2}), (\ref{xp1}), and we
reproduce the system on the cone. Because of the change of the
nature of the constraints from the second to the first class, the
described limit procedure has a rather formal character, however a
relationship between DSR transformations and the symmetries of a
AdS${}_{d+1}$ of infinite radius can be observed, because the DSR
particle in $d$ is isomorphic to one moving in AdS${}_{d+1}$ with
Lorentz invariance.

\section{Discussion and outlook}

To conclude, let us summarize the obtained results and discuss
shortly some problems that deserve a further attention.

In section $2$, the Conformal Group of the massless relativistic
particle was reviewed and in section $3$, it was obtained as a
dimensional reduction from a $\mathbb{R}^{d+2}$ space. In section
$4$, Because of the interchangeability of momenta and positions in
the Hamiltonian formulation, a new generator was added. With this
generator, the Conformal Group can be extended and a non linear
algebra rose.  In section $4$, starting with the formulation of
Magueijo Smolin deformed boosts, they are realized as a version of
the special conformal generator after a dimensional reduction onto
the mass shell of a massless relativistic particle and then, with
the Hamiltonian relations retrieved on that physical surface, the
velocity can be achieved from the transformed momenta.

Among the problems that deserve more attention is the problem of
how to include the mass term, that is still in darkness. On the
other hand, due the relation of DSR theories with AdS spaces
concluded here and in the bibliography, it could be interesting to
research about its relation with the AdS/CFT correspondence.
Finally to extend the studies to fields formulation could be a
interesting task too.

 \vskip 0.5cm {\bf Acknowledgements}
\vskip 5mm I want to thank to the Departamento de F\'{\i}sica de
la Universidad de Tarapac\'{a}.

\end{document}